\documentclass[aps,prb,twocolumn,groupedaddress,showpacs,floatfix]{revtex4}
\usepackage{graphicx}
\begin{document}
\pagenumbering{arabic} 
\pagestyle{plain}
\title{Superconducting anisotropy in the electron-doped high-T$_{c}$ superconductors Pr$_{2-x}$Ce$_{x}$CuO$_{4-y}$}
\author{Guoqing Wu,$^{1,2}$ R. L. Greene,$^{3}$ A. P. Reyes,$^{4}$ P. L. Kuhns,$^{4}$ W. G. Moulton,$^{4}$ Bing Wu,$^{5}$ Feng Wu, $^{6}$ and W. G. Clark $^{7}$}
\altaffiliation{Corresponding author electronic address: clark@physics.ucla.edu (W. G. Clark).}
\affiliation{$^{1}$College of Physics Science and Technology, Yangzhou University, Yangzhou, Jiangsu 225002 China}
\affiliation{$^{2}$Department of Physics, University of West Florida, Pensacola, Florida 32514, USA}
\affiliation{$^{3}$Department of Physics, University of Maryland, College Park, Maryland 20742, USA}
\affiliation{$^{4}$National High Magnetic Field Laboratory, Florida State University, Tallahassee, Florida 32306, USA}
\affiliation{$^{5}$Department of Math and Computer Science, Fayetteville State University, Fayetteville, NC 28301, USA}
\affiliation{$^{6}$School of Energy and Electrical Engineering, Hohai University, Nanjing, 210098, China}
\affiliation{$^{7}$Department of Physics and Astronomy, University of California, Los Angeles, California 90095, USA}
\date{\today}
\begin{abstract}
    We report superconducting anisotropy measurements in the electron-doped high-$T_{c}$ superconductors (HTSCs) Pr$_{2-x}$Ce$_{x}$CuO$_{4-y}$ (PCCO, $x$ = 0.15 and 0.17) with applied magnetic field ($H_{0}$) up to 28 T. Our results show that the upper critical field [$H_{c2}(T)$] is highly anisotropic, and as temperature $T$ $\rightarrow$ 0 the value of it at $H_{0}$ $\parallel$ $c$ [$H_{c2,\parallel c}$(0)] is far less than the Pauli limit, which is very different from that at $H_{0}$ $\perp$ $c$. The $H_{c2}(0)$ character along with the evaluated zero $T$ coherence length $[\xi_{ab(c)}(0)]$ and penetration depth $[\lambda_{ab(c)}(0)]$ is compared with those of the hole-doped cuprate HTSCs and typical Fe-based superconductors. We find that the low temperature anisotropic character of PCCO is rather similar to that of hole-doped cuprate HTSCs, but apparently larger than that of typical Fe-based superconductors. This study also proves a new sensitive probe of detecting rich properties of unconventional superconductors with the use of the resonant frequency of a NMR probe circuit.   
\end{abstract}
\maketitle
\section{Introduction}
%
    The determination of superconducting anisotropy with the values of upper critical field ($H_{c2}$), coherence length ($\xi$, $\sim$ vortex core diameter) and penetration depth ($\lambda$) that characterize the superconducting state of high-$T_{c}$ superconductors (HTSCs), including the electron-doped cuprate superconductors, is crucial for the understanding of high-$T_{c}$ superconductivity. The superconducting anisotropy coupled with short coherence length in hole-doped high-$T_{c}$ cuprates gives rise to a complex mean-field (MF) field-temperature ($H$ $-$ $T$) phase diagram, \cite{fisher, cyrot, blatter} which has an irreversibility field [$H_{m}(T)$] line falling below that of $H_{c2}(T)$, featuring a mixed superconducting state that includes a vortex liquid (reversible) and a vortex solid (lattice or glass, irreversible) with a penetrating magnetic field. These observations have been largely reported by the dc magnetic susceptibility measurements on hole-doped La$_{2-x}$Ba$_{x}$CuO$_{4}$, \cite{cyrot, muller} current-voltage (I$-$V) measurements on Y$_{1}$Ba$_{2}$Cu$_{3}$O$_{7-x}$, \cite{cyrot, worth} and magnetoresistance measurements on electron-doped cuprate $R_{1.85}$Ce$_{0.15}$CuO$_{4-x}$ ($R$ = Nd, Sm, Pr), \cite{herrmann, han, fournier} Fe-based superconductors K$_{x}$Fe$_{2}$Se$_{2}$, \cite{miz} LaFeAsO$_{x}$F$_{1-x}$ \cite{hunte} and many other type-II superconductors. \cite{solo, osofsky, gasp, zhang}
\begin{figure}
\includegraphics[scale= 0.27]{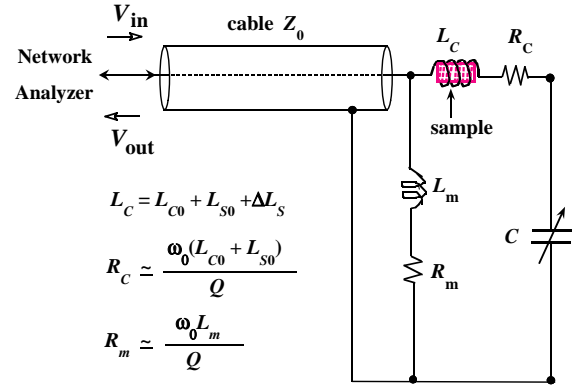}
\caption{(Color online) Sketch of the NMR probe circuit used for the superconducting anisotropy measurements on PCCO ($x$ = 0.15 and 0.17). The symbols and operation are described in the Section II text. \label{fig1}}
\end{figure}

     Recent studies show that the superconducting anisotropy in the hole-doped cuprates Y$_{1}$Ba$_{2}$Cu$_{3}$O$_{7-x}$ \cite{white, chang, ram} and La$_{2-x-y}$Eu$_{y}$Sr$_{x}$CuO$_{4}$ \cite{chang} is rather large and has a significant doping and/or in-plane field dependence, while the anisotropy in the Fe-based superconductors is generally smaller at low temperatures but has a rather strong temperature dependence. \cite{ved, yen} However, it is not clear for the electron-doped cuprate superconductors as an anisotroy of $\gamma$ = 15 [$\gamma$ $\equiv$ $H_{c2,\perp c}(0)/H_{c2,\parallel c}(0)$] obtained from the electron-doped cuprate Sr$_{1-x}$La$_{x}$CuO$_{2}$ \cite{jov} may not be as weak as one should claim on it.   
   
    On the other hand, controversy $\&$ inaccuracy exists \cite{zava} for the measurements of the superconducting anisotropy due to the limitations of the techniques used for the measurements, the requirement of high magnetic field, and the methods used to extract the values of $H_{c2,\perp c}(T)$ and $H_{c2,\parallel c}(T)$ [including $H_{c2,\perp c}(0)$ and $H_{c2,\parallel c}(0)$], etc.
 
    Conventional techniques for measuring the superconducting anisotropy use the electrical resistivity ($\rho$), \cite{cyrot, ando, fournier, suzuki} the imaginary part of the sample ac susceptibility ($\chi^{\prime\prime}$), \cite{fournier} or SQUID dc susceptibility ($\chi^{\prime}$). \cite{li} A problem with the resistivity measurements is that in HTSCs the resistive transition is usually quite broad. The extracted values of $H_{c2}$ and $T_{c}$ are actually the field and temperature relative to a fraction of the "normal-state" resistivity, respectively [see Supplemental Material]. \cite{see-SM}

    Thus the values obtained from the behavior of the electrical resistivity have a substantial uncertainty because they depend on the percentage for the onset of the flux flow used to determine them, while the criterion for the values obtained from the susceptibility measurements usually has a sensitivity issue that based on the noise level. \cite{herrmann, fournier, ando} Because the vortex solid state actually extends to a much broader range of field than what one would expect from the resistivity measurements, \cite{wang} novel measurement techniques are highly valuable, as indicated by recent Nernst effect measurements in both cuprate and Fe-based HTSCs. \cite{wang, zhu, hess1, hess2, ong, balci}

     In this paper, we report the superconducting anisotropy measurements on single crystals of the electron-doped HTSCs Pr$_{2-x}$Ce$_{x}$CuO$_{4-y}$ (PCCO, $x$ = 0.15 and 0.17) with applied magnetic field $H_{0}$ up to 28 T, using the resonant frequency ($f_{R}$) of a nuclear magnetic resonance (NMR) probe circuit, which can sensitively track the effect of the changes in the superconducting vortex phases on the shielding of the radio frequency magnetic field by the sample that is placed in the NMR coil. To our knowledge, this is the first report of this method being used to measure the superconducting anisotropy, and being able to combine both superconducting anisotropy and NMR measurements \cite{wu} in the same run is of added value.

     Our main results are that the upper critical field $H_{c2}$ of PCCO is highly anisotropic with significantly smaller values of $H_{c2,\perp c}$ than those of the hole-doped cuprate HTSCs, which indicates a significantly smaller vortex liquid regime for the electron-doped than for the hole-doped cuprate HTSCs. As $T$ $\rightarrow $ 0 the upper critical field [$H_{c2,\parallel c}$(0)] at $H_{0}$ $\parallel$ $c$, is far less than the Pauli limit $H_{\text{Pauli}}$(0). Thus, it is very different from that at $H_{0}$ $\perp $ $c$. Other anisotropies are the zero $T$ coherence length $[\xi_{ab(c)}(0)]$ and penetration depth $[\lambda_{ab(c)}(0)]$, which are also compared with those of hole-doped counterparts. Our results indicate that the low $T$ anisotropies appear rather similar for both electron-doped and hole-doped cuprate HTSCs, but apparently larger than those of typical Fe-based superconductors. Our measurements also prove a novel sensitive probe of detecting rich properties of unconventional superconductors with the use of the resonant frequency of a NMR probe circuit.   
\section{experimental details}

     Single crystals of PCCO ($x$ = 0.15 and 0.17) were grown with a flux technique. \cite{peng, brinkmann} The samples used in this experiment including the angular dependence of $\it{T}_{c}$ measurements are the same ones as we used for other NMR measurements. The sample size for PCCO ($x$ = 0.15) is $\sim$ 1.5 mm $\times$ 1.2 mm $\times$ 35 $\mu$m with a mass of 0.53 mg, and the size of the PCCO ($x$ = 0.17) sample is similar. The NMR coil (inductance $L_{\text{C}}$) was made from 50 $\mu$m silver wire wound with $\sim$ 20 turns. It has a quality factor ($Q$ $\sim$ 60) and is attached to a goniometer with the sample rotation axis that is $\perp$ $H_{0}$ and located in the lattice $ab$-plane. The high field measurements were conducted at the National High Magnetic Field Laboratory in Florida. A commercial network analyzer (NA) was used for detecting $f_{\text{R}}$. The NMR probe was built by W. G. Clark' group at UCLA. The measurements used a value of $f_{\text{R}}$ $\simeq$ 300 MHz for the major two alignments ($H_{0}$ $\parallel $ $c$ and $H_{0}$ $\perp $ $c$), with a resolution of 50 kHz or 0.017\%. For angular dependence measurements at $H_{0}$ = 9 T, $f_{\text{R}}$ $\sim$ 100 MHz was used, as we considered the convenience of our combined NMR spin-echo spectrum and spin-lattice relaxation measurements. \cite{wu} We do not expect frequency dependence in our $H_{c2}$ and $\it{T}_{c}$ measurements.

     Figure 1 shows a sketch of the NMR probe circuit used for the $H_{c2}$ measurements. The components are the NMR coil (inductance $L_{\text{C}}$), its resistance ($R_{\text{C}} = 2\pi f L_{\text{C}}/Q$), the series tuning capacitance ($C$), the parallel matching inductance ($L_{\text{m}}$), and its series resistance ($R_{\text{m}}$). To include the effects of the sample in the coil, we use $L_{\text{C}}=L_{\text{C0}}+L_{\text{S0}}+\Delta L_{\text{S}} = L_{\text{0}}+\Delta L_{\text{S}}$, where $L_{\text{C0}}$ is the inductance of the empty coil, $L_{\text{S0}}$ is the change in $L_{\text{C}}$ caused by the sample at the start of the measurement (frequency $f_{\text{0}}$), $L_{\text{0}} = L_{\text{C0}} + L_{\text{S0}}$ is the value of $L_{\text{C}}$ at the start of the measurement, and $\Delta L_{\text{S}}$ is the change in $L_{\text{C}}$ from the sample during the measurement. The real part of $\Delta L_{\text{S}}$ (Re$\{{{\Delta L_{\text{S}}}}\}$) can be + or $-$ (reduced or increased shielding). The imaginary part (Im$\{{\Delta L_{\text{S}}}\}$) is negative and represents the change in the rf losses associated with the dissipation from the shielding currents in the sample. 

     Scaling between $L_{\text{C0}}$ and the characteristic impedance of the cable ($Z_{\text{0}}$ = 50 $\Omega$) is given by 2$\pi f L_{\text{C0}} = k Z_{\text{0}}$. Typically, $0.5<k<2$ and $L_{\text{C0}} >> L_{\text{S0}}$, $\Delta L_{\text{S}}$. The coil circuit (reactance $Z_{\text{T}}$) is connected by a coaxial transmission line to the NA, which sends a signal (amplitude $V_{\text{in}}$, swept $f$) to the NMR coil circuit and receives the corresponding reflected signal ($V_{\text{out}}$). The values of $L_{\text{C}}$, $C$, and $L_{\text{m}}$ are chosen to make $Z_{\text{T}}\simeq Z_{\text{0}}$ close to the frequency ($f_{\text{0}}$) at which the measurement is started. It can be shown that at moderate and high values of $Q$, this is done by making 2$\pi f_{\text{0}}L_{\text{C0}}\simeq 1/(2\pi f_{\text{0}}C)$ and 2$\pi f_{\text{0}}L_{\text{m}}\simeq Z_{\text{0}}\sqrt{k/Q}$.
\begin{figure}
\includegraphics[scale= 0.27]{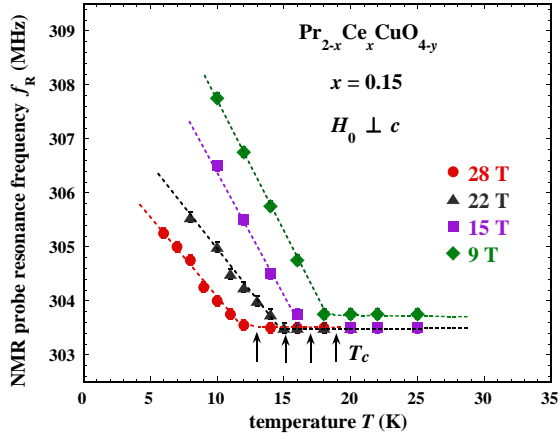}
\caption{(Color online) The measured NMR probe circuit resonance frequency $f_{\text{R}}$ vs $T$ for PCCO $x$ = 0.15 at $H_{0}$ $\perp$ $c$. The dashed lines are guides to the eye. \label{fig2}}
\end{figure}

     The NA measures the reflection coefficient $(\Gamma (f)=[Z_{\text{T}}(f) - Z_{\text{0}}]/[Z_{\text{T}}(f)+ Z_{\text{0}}])$ associated with $V_{\text{in}}$ and $V_{\text{out}}$ over a range of $f$. By using the conditions above to obtain $Z_{\text{T}}\simeq Z_{\text{0}}$ near $f_{\text{0}}$, a minimum in the magnitude of $\Gamma (f)$ occurs at the nearby frequency $f_{\text{R}}$. At the start of the measurement, a slight adjustment is made in $C$ to set $f_{\text{R}}=f_{\text{0}}$. The value of $f_{\text{R}}$ for each set of conditions is provided by the NA from the $f$-sweep of $\left \vert \Gamma (f)\right \vert$ displayed on its screen. Our numerical analysis of $\left \vert \Gamma(f) \right \vert$ for the circuit above shows that to a good approximation \cite{guoqing} and over a fairly large range of $Q$ and $k$, 
\begin{equation} 
f_{\text{R}}\simeq f_{0}\left(1 - \text{Re}{\{L_{\text{S}}}\}/2L_{0}\right). 
\end{equation}

     It also turns out that variations in Im$\{{L_{\text{S}}}\}$ and $Q$ have a significant effect on the depth of the minimum in $\left \vert \Gamma (f)\right \vert $, but have a negligible effect on $f_{\text{R}}$. As shown below, the value of $f_{\text{R}}$ near 300 MHz in our measurements can be shifted from $f_{0}$ up to $\sim $ $\pm$ 5 MHz, which is much larger than the accuracy and resolution of the readings using the NA, indicating that this method has significant value in comparison with other techniques.

     An important consideration for the sensitivity in these measurements is that the sample fills the coil as much as possible. In these measurements, the sample fills about 60\% of the volume of the NMR coil.
\section{experimental results}
     Figure 2 shows the measured values of $f_{\text{R}}$ vs $T$ for the single crystal of PCCO $x$ = 0.15 at several applied magnetic fields $H_{0}$ $\perp$ $c$ up to 28 T. Similar result for the sample PCCO $x$ = 0.17 at $H_{0}$ $\perp$ $c$ is shown in the Fig. 3. The result of $f_{\text{R}}$ vs $T$ for $H_{0}$ $\parallel$ $c$ is shown in Fig. 4, as an example for PCCO $x$ = 0.15. 

     As shown in Figs. 2$-$4, $f_{\text{R}}$ almost linearly increases with $T$ below $T_{c}$ where the flux-flow starts as indicated by the arrows (before $f_{\text{R}}$ reaches saturation). For PCCO $x$ = 0.15, the value of $f_{\text{R}}$ at $B_{0}$ $\perp$ $c$, $f_{R, \perp c}$ $\sim$ 303.5 MHz above $T_{c}$ (Fig. 2) which is in the normal state (note, no frequency deviations above temperatures $T$ $>$ $T_{c}$). Below $T_{c}$ it reaches $\sim$ 308.0 MHz by decreasing $T$ to $\sim$ 8 K at 9 T. Thus this indicates that the circuit resonant frequency has a shift of up to $\sim$ 4.5 MHz due to the change of the sample susceptibility (other effect that contributes to the shift is estimated to be very small since the range of temperature change is rather narrow).
\begin{figure}
\includegraphics[scale= 0.27]{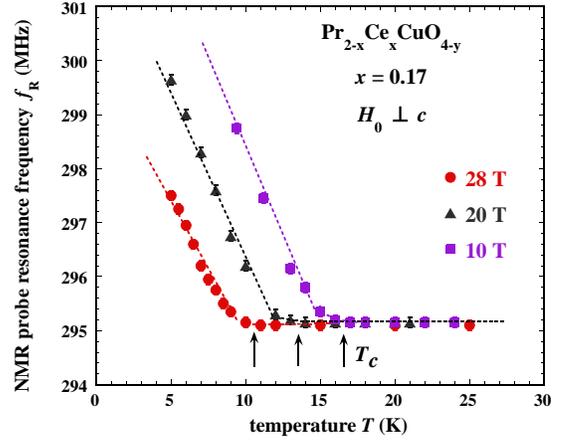}
\caption{(Color online) The measured NMR probe circuit resonance frequency $f_{\text{R}}$ vs $T$ for PCCO $x$ = 0.17 at $H_{0}$ $\perp$ $c$. The dashed lines are guides to the eye. \label{fig3}}
\end{figure}
\begin{figure}
\includegraphics[scale= 0.32]{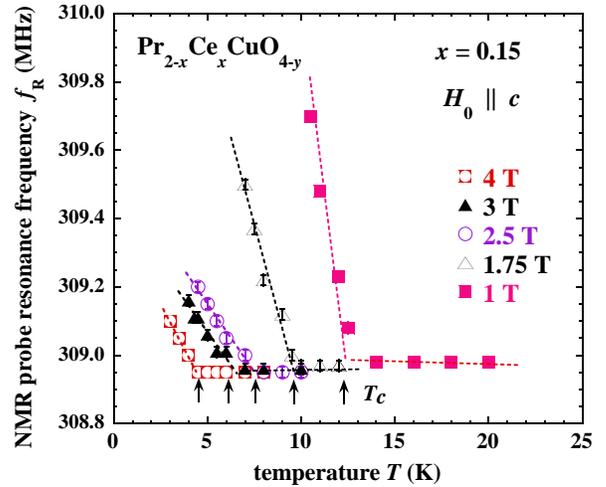}
\caption{(Color online) The measured NMR probe circuit resonance frequency $f_{\text{R}}$ vs $T$ for PCCO $x$ = 0.15 at $H_{0}$ $\parallel$ $c$. The dashed lines are guides to the eye. \label{fig4}}
\end{figure}

     Generally, the loss of density of states at the Fermi level, when the sample enters the flux lattice (glass) state, will cause $\text{Re}{\{L_{\text{S}}}\}$ to decrease. Thus it is expected that $f_{\text{R}}$ will go up on traversing from the normal to the SC state, as seen from Eq. (1) and by the data (Figs. 2$-$4). 

     Noticeably, at $H_{0}$ $\parallel$ $c$ the increase of $f_{\text{R}}$ below $T_{c}$ is much slower, and the applied magnetic field $H_{0}$ $\parallel$ $c$ corresponding to its $T_{c}$ is also significantly smaller than those at $H_{0}$ $\perp$ $c$. The anisotropy of the sample magnetic susceptibility should play a significant role for these differences. 

     In order to reveal this strong anisotropy effect to the $T_{c}$ and $f_{\text{R}}$, we performed the angular dependence of $T_{c}$ measurements by measuring the $f_{\text{R}}$ vs $T$ with various alignment angles ($\theta$) of the applied magnetic field for PCCO $x$ = 0.15 at $H_{0}$ = 9 T. These are shown in Fig. 5 and its inset, respectively. 

     As it is shown in Fig. 5, $T_{c}$ has a very strong angular dependence as a change in angle $\theta$ by 1.5$^{\circ}$ near $H_{0}$ $\perp$ $c$ will cause a change in $T_{c}$ by $\sim$ 2 K. On the other hand, upon cooling down to low enough temeratures the resonance frequency $f_{\text{R}}$ saturates [Fig. 5 inset] at $T_{m}$, the temperature corresponding to that where the electrical resistivity $\rho$ begins to be zero upon cooling in temperature, or the temperature corresponding to that where the electrical resistivity $\rho$ starts to deviate from zero while warming up  [see Supplemental Material]. \cite{see-SM}

     Figure 5 indicates a range of angle $\theta$ = $\sim$ $\pm$ 30$^\circ$ in PCCO $x$ = 0.15 that the applied magnetic field is not able to destroy its superconductivity completely. This range is rather wide, for the applied magnetic field alignment apart from  $H_{0}$ $\perp$ $c$ while superconductivity can still survive in PCCO ($x$ = 0.15) when $H_{0}$ = 9 T, as compared with many other two-dimensional superconductors including the field-induced two-dimensional organic superconductor $\lambda-$(BETS)$_{2}$FeCl$_{4}$ under similar conditions. \cite{ujj, balicas} This suggests that the electron-doped cuprate superconductors PCCO have relatively rather stable Cu 3d-electron orbitals in the Cu-O plane.
\begin{figure}
\includegraphics[scale= 0.32]{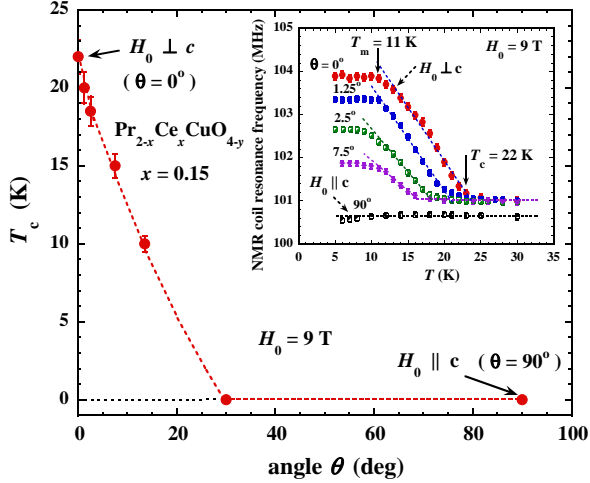}
\caption{(Color online) The measured angular dependence of $T_{C}$ and NMR probe circuit resonance frequency $f_{\text{R}}$ vs $T$ for various alignment angles ($\theta$) of the applied magnetic field for PCCO $x$ = 0.15 (inset) at $H_{0}$ = 9 T ($f_{\text{R}}$ is chosen to be $\sim$ 100 MHz without sample in the probe coil), where $\theta$ = 0$^\circ$ is for $H_{0}$ $\perp$ $c$, and $\theta$ = 90$^\circ$ is for $H_{0}$ $\parallel$ $c$. The dashed lines are guides to the eye. \label{fig5}}
\end{figure}
\begin{figure}
\includegraphics[scale= 0.32]{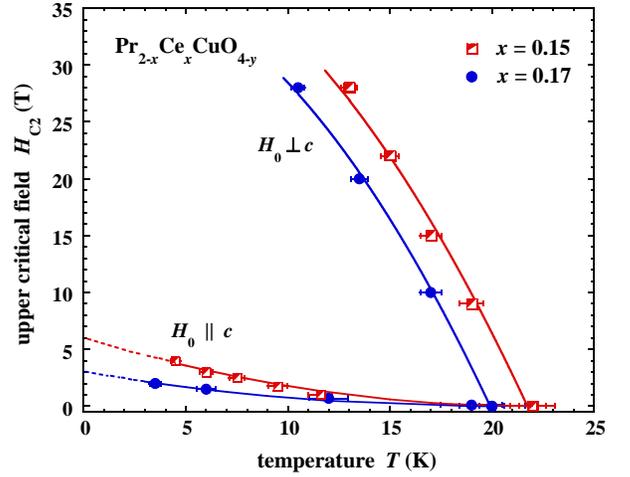}
\caption{(Color online) The irreversibility field line $H_{c2}$ vs $T$ for PCCO $x$ = 0.15 and 0.17 with $H_{0}$ $\parallel $ $c$ and $H_{0}$ $\perp $ $c$, obtained from the NMR probe circuit resonance frequency $f_{\text{R}}$ vs $T$ measurements. The values of $H_{c2}$ correspond to the values of $H_{0}$ with $T$ = $T_{c}$ as indicated by the arrows in the Figs. 2 - 4. The solid (dashed) lines are fits to Eq. (2) [$H_{0}$ $\parallel $ $c$] and Eq. (4) [$H_{0}$ $\perp$ $c$]. \label{fig6}}
\end{figure}
\section{Discussion}
     Figure 6 shows the values of the upper critical field $H_{c2}$ vs $T$ obtained from the data as shown in Figs. 2$-$4, in which the values of $H_{c2}$ corresponds to the values of $H_{0}$ with $T$ = $T_{c}$ as indicated by the arrows in the figures. At low temperatures, the values of $H_{c2, \perp c}$ are significantly larger than those of $H_{c2, \parallel c}$, indicating a high anisotropy character.

     At $H_{0}$ $\parallel$ $c$, the $H_{c2, \parallel c}$ vs $T$ curve has an anomalous shape (a concave upward curvature), which has been widely observed in both hole- and electron-doped cuprate HTSCs \cite{osofsky, ando, welp, zheng1, dali, hidaka, wang, han, fournier, herrmann} and many other systems as well, where there is no evidence of saturation as $T$ $\rightarrow$ 0. However, the mechanism is unknown, especially as $T$ $\rightarrow$ 0. More peculiarly, hole-doped cuprate HTSCs even show \cite{ando, osofsky} a divergence of $H_{c2, \parallel c}(T)$ at low $T$. 

     However, for the electron-doped cuprate HTSCs, it is generally agreed \cite{ong, herrmann} that as $T$ $\rightarrow $ 0 the upper critical field $H_{c2, \parallel c}$($T$ $\rightarrow$ 0) = $H_{c2, \parallel c}$(0) = $\sim$ $H_{m, \parallel c}$(0). Considering this, one can fit the data more appropriately as \cite{escote} 
\begin{equation}
H_{c2, \parallel c}(T) \approx H_{c2, \parallel c}(0)~ [1 - T/T_{c}]^{\alpha},
\end{equation}
rather than using the Werthamer, Helfand, and Hobenberg (WHH) formula \cite{whh}
\begin{equation}
H_{c2}(0) = - 0.693T_{c}(dH_{c2}/dT)_{T_{c}},
\end{equation}
where Eq. (2) is consistent \cite{escote} with the Ginszburg-Landau theory. \cite{cyrot} 

     Here in Eq. (2), $T_{c}$ = 22 K and 20 K for $x$ = 0.15 and 0.17, respectively. This gives a result of $\alpha$ $\approx$ 2.0, and $H_{c2, \parallel c}$(0) = (6.0 $\pm$ 0.3) T and (3.1 $\pm$ 0.4) T for $x$ = 0.15 and 0.17, respectively, which agrees well with the result from the field dependence of the specific heat measurements. \cite{balci1} Similar results were also obtained from the electrical resistivity measurements \cite{herrmann} on Nd$_{2-x}$Ce$_{x}$CuO$_{4-y}$ with $\alpha$ = 1.7 - 2.0. 
\begin{table*}
\caption{A comparison of fit parameters of PCCO ($x$ = 0.15 and 0.17) with those of hole-doped cuprate HTSCs and typical Fe-based superconductors. 1 $-$ ref. [2], 2 $-$ result of these measurements, 3 $-$ result of resistivity measurements (refs. [8, 52]) from our analysis, 4 $-$ ref. [53-56], 5 $-$ ref. [57-58], 6 $-$ ref. [59-60], and 7 $-$ ref. [61]. \label{tab1}}
\begin{ruledtabular}
\begin{tabular}{ccccccc}
& & & Upper critical field (T) & Coherence length $(\rm\AA)$ & Penetration depth $(\rm\AA)$ \\ \hline 
& Compound & $T_{c}$ (K) & $H_{c2,\perp c}(0)$/$H_{c2,\parallel c}(0)$ & $\xi_{ab}(0)$/$\xi_{c}(0)$ & $\lambda_{ab}(0)$/$\lambda_{c}(0)$ \\ \hline
& $^{1}$La$_{2-x}$Sr$_{x}$CuO$_{4-y}$ & 38 & 80/15 & 35/7 & 800/4000 \\ \hline
& $^{1}$YBa$_2$Cu$_3$O$_7$ & 92 & 150/40 & 15/4 & 1500/6000 \\ \hline
& $^{1}$Bi$_2$Sr$_2$Ca$_2$Cu$_3$O$_{10}$ & 110 & 250/30 & 13/2 & 2000/10000 \\ \hline
& PCCO ($x$ = 0.15) & 22$^{2,~3}$ & 42/6$^{2}$, ~~~ 55/5.5$^{2,~3}$ & 74/10$^{2}$ & 2500/18500$^{2}$ \\ \hline
& PCCO ($x$ = 0.17) & 20$^{2,~3}$ & 38/3$^{2}$, ~~~ 40/3$^{2,~3}$ & 103/8$^{2}$ & ............... \\ \hline
& $^{4}$LiFeAs            & 18 & 24/15 & 48/17  & ............... \\ \hline
& $^{5}$Fe$_{1.1}$Se$_{0.6}$Te$_{0.4}$ & 14 & 47/47 & 26.5/26.5  & ............... \\ \hline
& $^{6}$(Ba,K)Fe$_{2}$As$_{2}$ & 28 & 57/55$^{*}$ & 21.7/21.7  & ............... \\ \hline
& $^{7}$SmFeAsO$_{0.85}$ & 50 &  51$^{**}$/56$^{***}$  & 170/36  & ............... \\ 
\end{tabular}  \footnote{$^{*}$ values at $\sim$ 10 K, $^{**}$ 43 K, and $^{***}$ 27 K.} 
\end{ruledtabular}
\end{table*}

     The property that $\alpha$ $\approx$ 2.0 for PCCO instead of $\alpha$ = 1 suggests that there exists a significant distribution of inhomogeneous vortex state and local SC regions, \cite{caix} which could serve \cite{fournier, fisher} as a possible mechanism leading to the vortex-glass phase as that observed in Nd$_{2-x}$Ce$_{x}$CuO$_{4-y}$. \cite{herrmann, wang1}

     At $H_{0}$ $\perp$ $c$, the data of $H_{c2, \perp c}$ vs $T$ ($T/T_{c}$ $>$ 0.5) shown in Fig. 6 can be fitted with a BCS $T-$dependence (also used by Clem \cite{clem} earlier) for $H_{c2, \perp c}$. Since it is also known \cite{ong, herrmann} that $H_{c2, \perp c}$(0) = $H_{c2, \perp c}$($T$ $\rightarrow$ 0) = $\sim$ $H_{m, \perp c}(0)$ as $T$ $\rightarrow$ 0, it is appropriate to use \cite{clem}
\begin{equation}
H_{c2, \perp c}(T) \approx H_{c2, \perp c}(0)~ [1 - (T/T_{c})^{2}].
\end{equation}
This gives a fitted value of $H_{c2, \perp c}(0)$ $\sim$ 42 T and 38 T for $x$ = 0.15 and 0.17, respectively, by the extrapolation of Eq. (4) to $T$ = 0, as shown by the dark solid and dashed lines in Fig. 6.

     According to the BCS theory, \cite{bcs} the energy gap $\Delta$(0) (SC gap) between the normal state and the SC state at $T$ = 0 is related to the Pauli limit field $H_{\text{Pauli}}(0)$ as \cite{clogston} 
\begin{equation}
\Delta(0) = \sqrt{2}\mu_{B} H_{\text{Pauli}}(0),
\end{equation}
where $\mu_{B}$ is the Bohr magneton. The result from the tunneling measurements \cite{biswas1} indicates that at $H_{0}$ $\parallel$ $c$, the value of 2$\Delta(0)$ $\approx$ 4.0 $k_{B}T_{c}$ and 3.4 $k_{B}T_{c}$ for PCCO $x$ = 0.15 and $x$ = 0.17, respectively. This gives the Pauli limit $H_{\text{Pauli}}(0)$ $\sim$ 45 T and $\sim$ 36 T for $x$ = 0.15 and 0.17.

     Thus, $H_{c2, \parallel c}(0)$ is far less than the Pauli limit $H_{\text{Pauli}}(0)$, while $H_{c2, \perp c}(0)$ is close to it, i.e., $H_{c2, \perp c}(0)$ $\approx$ $H_{\text{Pauli}}(0)$, and $H_{c2, \parallel c}(0)$ $<<$ $H_{\text{Pauli}}(0)$.

     To estimate the SC coherence length at $T$ = 0, one can use the expression \cite{cyrot} $H_{c2,\parallel c}(0)$ = $\phi _{0}$/2$\pi \xi _{ab}^{2}(0)$, and $H_{c2,\perp c}(0)$ = $\phi _{0}$/2$\pi \xi _{ab}(0)\xi _{c}(0)$, where the flux quantum $\phi_{0}$ = 2.07 $\times $ 10$^{-7}$ G.cm$^{2}$, and $\xi_{ab}(0)$ and $\xi _{c}(0)$ are the CuO$_{2}$ in-plane and out-of-plane coherence length at $T$ = 0, respectively. These yield $\xi _{ab}(0)$ $\sim$ 74 $\rm{{\mathring{A}}}$ and $\sim $ 103 $\rm{{\mathring{A}}}$ for PCCO $x$ = 0.15 and 0.17, respectively. Correspondingly, their out-of-plane values are $\xi _{c}(0)$ $\sim $ 10 $\rm{{\mathring{A}}}$ and $\sim$ 8 $\mathrm{{\mathring{A}}}$, respectively, which are slightly larger than the distance between the nearest CuO$_{2}$ planes. 

     Considering the in-plane penetration depth $\lambda_{ab}(0)$ $\sim$ 2500 $\rm{{\mathring{A}}}$ from the microwave measurement \cite{prozorov} for PCCO $x$ = 0.15, and using the anisotropy relationship \cite{cyrot} ($\varepsilon$ $\equiv$ $\xi_{c}/\xi_{ab}$ = $\lambda_{ab}/\lambda_{c}$), we have $\lambda_{c}(0)$ $\sim$ 18500 $\rm{{\mathring{A}}}$ for PCCO $x$ = 0.15. The anisotropy field ratio $H_{c2, \perp c}(0)$/$H_{c2, \parallel c}(0)$ $\sim$ 7 for $x$ = 0.15 and $\sim$ 12 for PCCO $x$ = 0.17.   

     A comparison with typical hole-doped cuprate HTSCs and Fe-based superconductors in the literature is listed in Table I. As shown in Table I, in comparison with typical hole-doped cuprate HTSCs the upper critical field $H_{c2}(0)$ values for PCCO are a few times smaller, and the coherence length $\xi(0)$ and penetration depth $\lambda(0)$ are a few time larger. But their low temperature anisotropies (values of $H_{c2,\perp c}(0)$/$H_{c2,\parallel c}(0)$ $\equiv$ $\gamma$) are rather similar.

     However, in comparison with the typical Fe-based superconductors, the anisotropies (at low temperatures) in the Fe-based superconductors are apparently smaller. The mechanism for their difference in anisotropies is not understood currently in the literature.  
    
     Here we would like to point out that with the same analysis method that we used here the difference between the obtained value of $H_{c2, \perp c}(0)$ in this experiment (UCLA sample) and that reported by the electrical resistivity measurements on a different sample (UM sample) \cite{li2} for PCCO $x$ = 0.17 is negligible, while for PCCO $x$ = 0.15 there is a rather significant difference for the corresponding values between the two samples. For clarity, we plotted the data together for both samples, which are shown in Fig. 7. The corresponding data of $H_{m}(T)$ for the UM sample are the ones that labeled as $H_{ext}$ in refs. [8] and [52] for $H_{0}$ $\parallel $ $c$ and $H_{0}$ $\perp $ $c$, respectively.
\begin{figure}
\includegraphics[scale= 0.32]{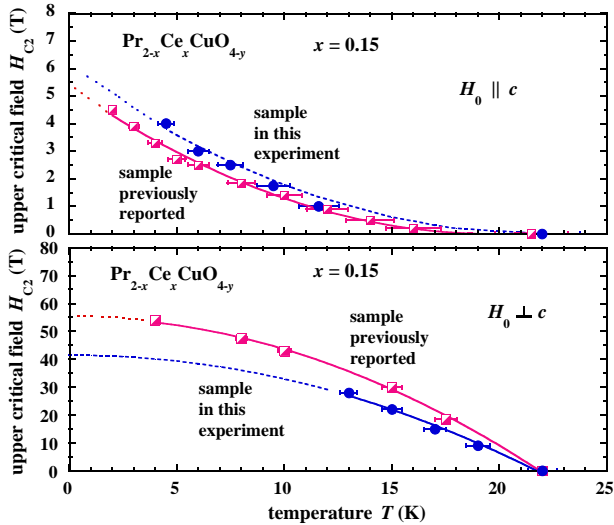}
\caption{(Color online) Comparison of obtained values of $H_{c2}$ vs $T$ data between the sample used in this experiment (blue curve) and a different one reported previously (red curve) from the electrical resistivity measurements \cite{fournier, li2} for PCCO $x$ = 0.15, with the same analysis method used here. The solid and dashed lines are the fit and the extrapolations (see text). \label{fig7}}
\end{figure}

     The corresponding values of $H_{c2, \perp c}(0)$ and $H_{c2, \parallel c}(0)$ obtained here for the UM sample are $\sim$ 55 T and $\sim$ 5.5 T, respectively, while the results for the UCLA sample from this experiment described above are $\sim$ 42 T and $\sim$ 6.0 T, respectively, i.e. the difference is $\sim$ 20$\%$. One possible cause for the difference could be due to the difference in the measurement method and/or in sample thickness.  
\section{conclusions}
     In summary, we reported the superconducting anisotropy measurements on the electron-doped high-$T_{c}$ superconductors PCCO ($x$ = 0.15 and 0.17) with the applied magnetic field $H_{0}$ up to 28 T, by the resonant frequency $f_{R}$ of the NMR probe circuit. The frequency data showed very sharp features at the phase transition temperature $T_{c}$ and almost a linear relation between $f_{R}$ and $T$ below $T_{c}$ (a sharp increase of $f_{R}$), thus indicating a significant advantage over the method using the electrical resistivity for the determination of $T_{c}$ and $H_{c2}$ values. 

     The measured values of $H_{c2}$ are highly anisotropic, and the analysis indicates that $H_{c2, \parallel c}$(0) is far less than the Pauli limit $H_{\text{Pauli}}(0)$, thus it is very different from that at $H_{0}$ $\perp$ $c$. A phase diagram that involves vortex solid/liquid states depending on the alignment of $H_{0}$ relative to the $c$-axis is proposed, and the obtained anisotropic $H_{c2}(0)$ character along with the evaluated coherence length $\xi_{ab(c)}(0)$ and penetration depth $\lambda_{ab(c)}(0)$ (at $T$ = 0) is compared with those of hole-doped cuprate HTSCs and the recently found Fe-based superconductors. The low temperature upper critical field $H_{c2}(0)$ values for PCCO are a few times smaller than those of the hole-doped cuprates, but their anisotropies are rather similar, while the typical Fe-based superconductors apparently have smaller anisotropies than PCCO at low temperatures. 

     The work at NHMFL was supported by NSF under Cooperative Agreement No. DMR-0654118 and the State of Florida, at UM by DMR-1104256 (R.L.G.), at University of West Florida by SCA/2009-2012 (Guoqing Wu), and at UCLA  by NSF Grants DMR-0334869 (W.G.C.). We thank Stuart E. Brown for helpful discussions and support. 

\end{document}